# La$_{1-x}$Bi$_{1+x}$S$_3$ (x~0.08): An *n*-Type Semiconductor

Fei Han,[†] Huimei Liu,[‡] Christos D. Malliakas,[†,§] Mihai Sturza,[†] Duck Young Chung,[†] Xiangang Wan,[‡] and Mercouri G. Kanatzidis[*,†,§]

[†] Materials Science Division, Argonne National Laboratory, Argonne, Illinois 60439, United States

[‡] National Laboratory of Solid State Microstructures, School of Physics, Collaborative Innovation Center of Advanced Microstructures, Nanjing University, Nanjing 210093, China

[§] Department of Chemistry, Northwestern University, Evanston, Illinois 60208, United States

**ABSTRACT:** The new bismuth chalcogenide La$_{0.92}$Bi$_{1.08}$S$_3$ crystallizes in the monoclinic space group $C2/m$ with $a$ = 28.0447(19) Å, $b$ = 4.0722(2) Å, $c$ = 14.7350(9) Å, and $\beta$ = 118.493(5)°. The structure of La$_{0.92}$Bi$_{1.08}$S$_3$ is built up of NaCl-type Bi$_2$S$_5$ blocks, and BiS$_4$ and LaS$_5$ infinitely long chains forming a compact three-dimensional framework with parallel tunnels. Optical spectroscopy and resistivity measurements reveal a semiconducting behavior with a band gap of ~ 1 eV and activation energy for transport of 0.36(1) eV. Thermopower measurements suggest the majority carriers of La$_{0.92}$Bi$_{1.08}$S$_3$ are electrons. Heat capacity measurements indicate no phase transitions from 2 to 300 K. Band structure calculations at the density functional theory level confirm the semiconducting nature and the indirect gap of La$_{0.92}$Bi$_{1.08}$S$_3$.

## INTRODUCTION

Antimony or bismuth chalcogenides have been broadly investigated as promising thermoelectric materials,[1-6] and many new type of compounds such as BaBiTe$_3$,[7] K$_2$Bi$_8$Se$_{13}$,[8] and CsBi$_4$Te$_6$[9,10] have been reported to possess novel properties. In 2009, the discovery of the single Dirac cone on the surface of Bi$_2$Se$_3$, Bi$_2$Te$_3$ and Sb$_2$Te$_3$ crystals further highlighted the importance of antimony or bismuth-chalcogenides for investigations as topological insulators.[11,12] After this, exploration of new phases was brought to the forefront again as part of a search for new topological insulators.[13-15]

Generally, antimony or bismuth-chalcogenides have narrow-band-gap or semi-metallic electronic structures.[16-18] At the boundary of semiconductor and metal, the electronic structures of antimony or bismuth-chalcogenides are varied from different systems or upon doping which is reflected in the tunability of thermoelectric performance and Fermi level of topological insulators to go across the Dirac cone.

Recently, superconductivity in bismuth-chalcogenides such as Cu$_x$Bi$_2$Se$_3$,[19] $Ln$O$_{1-x}$F$_x$BiS$_2$ ($Ln$ = rear earth),[20-22] CsBi$_4$Te$_6$[23] was discovered at the edge of semiconducting behavior. Among these bismuth-chalcogenide superconductors, $Ln$O$_{1-x}$F$_x$BiS$_2$ ($Ln$ = rare earth) attracted great attention due to its layered structure, unconventional nature of the material, and the implication of a possible family of BiS$_2$-based superconductors just like cuprates[24-26] and iron-based superconductors.[27,28] We first observed La$_{0.92}$Bi$_{1.08}$S$_3$ in a synthesis attempt targeting LaBiS$_3$, an analogue of LaOBiS$_2$ by replacing the PbO-type La$_2$O$_2$ layers with La$_2$S$_2$ layers. This chemistry, however, leads to La$_{0.92}$Bi$_{1.08}$S$_3$ which forms a quite different structure than LaOBiS$_2$. After our synthesis and structure determination, we noticed the stoichiometric compound LaBiS$_3$ has been reported in previous phase diagram studies,[29-31] but research on its structure and properties is lacking.

Here we describe the synthesis, the new type of crystal structure, detailed physical property measurements and electronic structure of La$_{0.92}$Bi$_{1.08}$S$_3$. The compound has a phase width (La$_{1-x}$Bi$_{1+x}$S$_3$) and is an *n*-type semiconductor with a band gap of ~ 1 eV.

## EXPERIMENTAL SECTION

**Sample Preparation.** Crystalline La$_{0.92}$Bi$_{1.08}$S$_3$ was synthesized by a solid state reaction. Precursors La$_2$S$_3$ and Bi$_2$S$_3$ were prepared by direct stoichiometric combination of La+S and Bi+S at 500 °C and 650 °C respectively with materials sealed in an evacuated silica tube. We caution that the reaction of La+S is highly exothermic and may lead to a tube explosion when the heating process is too fast. To avoid explosion, we used La scraps filed from a large La chunk instead of using commercial La powders and we heated the La+S mixture slowly at a rate of 10 °C/h to obtain La$_2$S$_3$. La$_2$S$_3$ and Bi$_2$S$_3$ in equimolar amounts were weighted, mixed and ground. The mixture was then loaded into an alumina crucible and sealed in an evacuated silica tube. All handling of chemicals was performed in a MBRAUN glove box under argon atmosphere (both H$_2$O and O$_2$ are limited below 0.1 ppm). The silica tube was then heated to and kept at 1000 °C for 10 h. Slow cooling from 1000 °C to 700 °C was carried out over 60 h to grow crystals. The resulting product contains small rod-shaped crystals on the top of the cylinder-like ingot. The crystals were determined La$_{0.92}$Bi$_{1.08}$S$_3$ in composition later by single-crystal X-ray diffraction. It is not clear from the size and morphology of the crystals if the crystal growth process is solid-state grain growth or assisted by a liquid or vapor phase. To grow bigger crystals, reactions with temperatures higher than 1000 °C were also attempted. However, above 1000 °C La$_{0.92}$Bi$_{1.08}$S$_3$ was found to decompose since a lot of Bi$_2$S$_3$ crystals could be found on the wall of crucibles. The compositions of LaBiSSe$_2$, LaBiSe$_3$, CeBiS$_3$, PrBiS$_3$, NdBiS$_3$, GdBiS$_3$, and HoBiS$_3$ were also examined but our reactions produced phase mixtures of binaries as determined by powder X-ray diffraction analysis. Efforts of making La$_{1-x}$Sr$_x$BiS$_3$ were unsuccessful presumably because of the high stability of SrS.

Morphology and elemental analyses of the La$_{0.92}$Bi$_{1.08}$S$_3$ crystals were performed on a Hitachi S-4700-II Scanning Electron Microscope with Energy Dispersive Spectroscope (EDS) detector equipped. The electron micrograph of La$_{0.92}$Bi$_{1.08}$S$_3$ in the inset of Figure 1 shows a rod-like crystal habit. Typical crystal dimensions are around 200 × 30 × 10 μm$^3$. Semiquantitative analysis by EDS indicates the crystal consists of La, Bi and S, as shown in Figure 1, with an average composition of La$_{0.204}$Bi$_{0.206}$S$_{0.590}$.

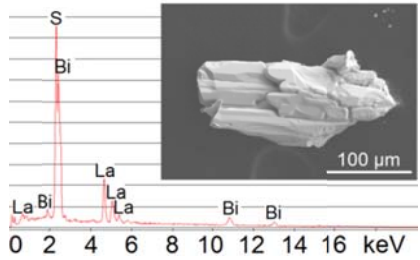

**Figure 1.** Energy Dispersive Spectrum (EDS) collected on a typical La$_{0.92}$Bi$_{1.08}$S$_3$ crystal whose electron micrograph is shown in the inset.

**Single-crystal X-ray Diffraction.** Single-crystal X-ray diffraction at room temperature was carried out on a STOE IPDS 2T diffractometer. Data reducing and absorption correction were performed with the software X-Area,[32] and the structure was solved by direct methods and refined using the SHELXTL software.[33] The resulting structure parameters are listed shortly in Table 1, 2 and in more detail in Table S1, S2, S3, and S4.

**Table 1. Room Temperature Single Crystal Data and Structure Refinement for La$_{0.92}$Bi$_{1.08}$S$_3$.**

| Empirical formula, weight | La$_{0.92}$Bi$_{1.08}$S$_3$, 449.97 |
|---|---|
| Wavelength | 0.71073 Å |
| Space group | Monoclinic, C2/m |
| Unit cell dimensions | a = 28.0447(19) Å, α = 90.00° |
| | b = 4.0722(2) Å, β = 118.493(5)° |
| | c = 14.7350(9) Å, γ = 90.00° |
| Volume, Z | 1478.96(15) Å$^3$, 12 |
| Density (calculated) | 6.063 g/cm$^3$ |
| Absorption coefficient | 47.553 mm$^{-1}$ |
| θ range for data collection | 4.19 to 34.78° |
| Index ranges | -44<=h<=44, -6<=k<=6, -23<=l<=23 |
| Reflections collected | 19673 |
| Independent reflections | 3487 [R$_{int}$ = 0.0778] |
| Completeness to θ = 24.97° | 97.3% |
| Data / restraints / parameters | 3487 / 0 / 100 |
| Goodness-of-fit | 0.926 |
| Final R indices [>2σ(I)] | R$_{obs}$ = 0.0319, wR$_{obs}$ = 0.0546 |

$R=\Sigma||F_o|-|F_c||/\Sigma|F_o|$, $wR=\{\Sigma[w(|F_o|^2-|F_c|^2)^2]/\Sigma[w(|F_o|^4)]\}^{1/2}$ and $w=1/[\sigma^2(F_o^2)+(0.0257P)^2]$ where $P=(F_o^2+2F_c^2)/3$

**Powder X-ray Diffraction.** The purity of the sample was checked with powder X-ray diffraction. Finely ground powders were mounted on a flat plate sample holder and were measured using a PAnalytical X'pert Pro diffractometer with an iron filtered Cu-Kα source, operating at 45 kV and 40 mA. The data were collected under a continuous scanning method in 2θ range of 10–80° using a step size of 0.0167°. Powder X-ray diffraction (XRD) data were analyzed with the Rietveld method using EXPGUI, the graphical interface of GSAS program.[34,35] It is found the crystals collected on the top of the ingot sample were pure while the main ingot contains minor La$_2$S$_3$ impurity. The powder XRD pattern for the ground crystals and its Rietveld fit are shown in Figure S1 in the supporting information. The data are well fitted with the C2/m structure solved by single-crystal X-ray diffraction. The Rietveld quality factors $R_p$ and $wR_p$ are 0.0292 and 0.0383 respectively, which indicates good agreement between the experimental and simulated data. To make property characterizations intrinsic, all the measurements were carried out on crystals or ground crystals.

**Table 2. Atomic Coordinates (×10$^4$) and Equivalent Isotropic Displacement Parameters (Å$^2$×10$^3$) for La$_{0.92}$Bi$_{1.08}$S$_3$ at Room Temperature. All sites are fully occupied.**

| Atom | x | y | z | Occupancy | U$_{eq}$* |
|---|---|---|---|---|---|
| Bi(1) | 1279(1) | 0 | 2852(1) | 1 | 16(1) |
| Bi(2) | 1327(1) | 0 | 7640(1) | 1 | 14(1) |
| Bi(3) | 2164(1) | 0 | 816(1) | 1 | 16(1) |
| La(1) | 2955(1) | 0 | 4088(2) | 0.747(7) | 11(1) |
| Bi(4) | 2876(2) | 0 | 4372(3) | 0.253(7) | 21(1) |
| La(2) | 4444(1) | 0 | 593(1) | 1 | 10(1) |
| La(3) | 5282(1) | 0 | 3959(1) | 1 | 11(1) |
| S(1) | 368(1) | 0 | 7727(2) | 1 | 11(1) |
| S(2) | 1218(1) | 0 | 4616(2) | 1 | 11(1) |
| S(3) | 1276(1) | 0 | 864(2) | 1 | 13(1) |
| S(4) | 2342(1) | 0 | 7634(2) | 1 | 17(1) |
| S(5) | 2861(1) | 0 | 6035(2) | 1 | 16(1) |
| S(6) | 3323(2) | 0 | 665(2) | 1 | 17(1) |
| S(7) | 3946(2) | 0 | 3654(2) | 1 | 23(1) |
| S(8) | 5590(1) | 0 | 2291(2) | 1 | 14(1) |
| S(9) | 0 | 0 | 5000 | 1 | 17(1) |
| S(10) | 0 | 0 | 0 | 1 | 15(1) |

*$U_{eq}$ is defined as one third of the trace of the orthogonalized $U_{ij}$ tensor.

**Heat Capacity.** We employed the thermal relaxation technique[36] to perform the heat capacity measurements on Quantum Design physical property measurement system (PPMS). To enhance the sample coupling, $La_{0.92}Bi_{1.08}S_3$ crystals were ground, pressed into a pellet, and annealed at 500 °C for 20 h. The compact round pellet was cut and milled into a square pellet which is a little bit smaller than the sample stage of heat capacity puck. N-grease was used as heat-conducting medium.

**Optical Spectroscopy.** The optical energy band gap of $La_{0.92}Bi_{1.08}S_3$ was determined using solid-state UV-Vis optical spectroscopy. The optical diffuse reflectance measurement at room temperature was performed using a Shimadzu Model UV-3101PC double-beam, double-monochromator spectrophotometer. $BaSO_4$ was used as a 100% reflectance standard. The absorption data were calculated based on the obtained reflectance spectra using the Kubelka-Munk equation as described elsewhere.[37]

**Resistivity and Thermopower.** Temperature-dependent resistivity measurements were also performed on a piece of crystal with PPMS. The resistivity by four-probe method was measured using 14 μm gold wire and silver paste to make electrical contacts. The resistance data was collected only between 320 K and 400 K because below 320 K the resistance was out of the measuring range of our PPMS.

To measure the thermopower of $La_{0.92}Bi_{1.08}S_3$, the pellet sample used in the heat capacity measurement was mounted on a heating stage and connected to the thermocouples with silver paste. The whole setup was evacuated in a chamber to keep the sample stable at high temperatures. The Seebeck voltage $V(T)$ was measured by the integral method, in which one end of the sample is held at a fixed temperature $T_0$, and the other end is varied through the temperature $T$ range of interest using a commercial MMR Technologies SB-100 Seebeck measurement system. The Seebeck coefficient $S$ is obtained from the slope of the $V(T)$ versus $T$ curve.[38]

**Electronic Structure Calculations.** The electronic band structure calculations were carried out using the full potential linearized augmented plane wave method as implemented in WIEN2K package.[39] The modified Becke-Johnson exchange potential together with local-density approximation for the correlation potential (MBJLDA) has been used here to obtain accurate band structures.[40] A 12×12×3 mesh is used for the Brillouin zone integral. Using the second-order variational procedure, we included the spin-orbital coupling (SOC) interaction in our calculations.

## RESULTS AND DISSCUSSION

**Crystal Structure Description.** The structure of $La_{0.92}Bi_{1.08}S_3$ projected onto the $ac$-plane is depicted in Figure 2(a). $La_{0.92}Bi_{1.08}S_3$ crystallizes in the space group $C2/m$ and presents a new type of three-dimensional structure. The structure is built up of $Bi_2S_5$ blocks, $BiS_4$ chains and $LaS_5$ chains which all extend infinitely along the $b$-axis and are interconnected in the $ac$-plane via sulfur atoms. Therefore, the overall structure is three-dimensional. An occupancy disorder of La(1) and Bi(4) atoms is present in the structure. They split in position due to different coordination preferences with the $S^{2-}$ anions. The positional splitting is clearly observed in the electron density distribution map generated by Fourier transform of the diffraction data, Figure 2(b).

The "$Bi_2S_5$" block can be regarded as a fragment excised out of the NaCl lattice. This $Bi_2S_5$ block consists of four $BiS_6$ octahedra in the $ac$-plane which are infinitely replicated in the $b$-direction, see Figure 3. NaCl-type $BiQ$ ($Q$ = Te, Se, S) structural units are a common building block in group 15 metal chalcogenides such as $Bi_2Te_3$, $KBi_{6.33}S_{10}$,[41] $K_2Bi_8Q_{13}$ ($Q$ = S,[41] Se[8,42]), $K_{2.5}Bi_{8.5}Se_{14}$,[8] $KM_4Bi_7Se_{15}$ ($M$ = Sn, Pb),[43] and $CsBi_4Te_6$.[9,10] In the $Bi_2S_5$ block of $La_{0.92}Bi_{1.08}S_3$, Bi(2) and Bi(3) are octahedrally coordinated by six $S^{2-}$ anions with bonding distances ranging from 2.524(2) to 3.3635(32) Å.

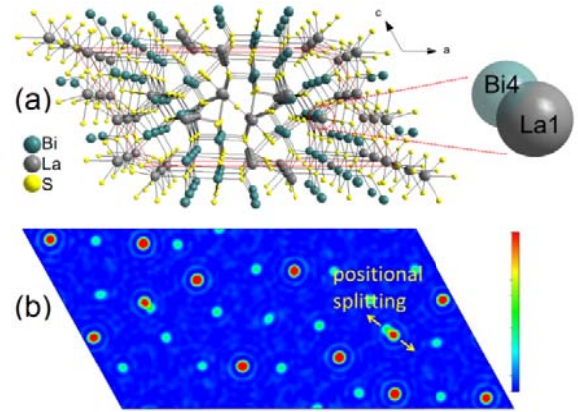

**Figure 2.** (a) Perspective view of the structure of $La_{0.92}Bi_{1.08}S_3$ along the crystallographic $b$-axis. (b) Electron density distribution sliced at $y = 1/2$ plane.

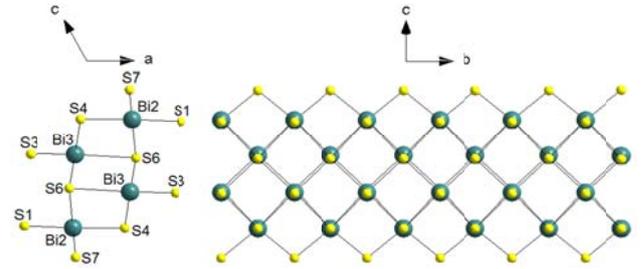

**Figure 3.** The one-dimensional $Bi_2S_5$ fragment with the NaCl-type lattice viewed along the $b$-axis and the $a$-axis respectively.

On the other hand, Bi(1) is in a crystallographically independent site in the structure. Bi(1) is also octahedrally coordinated by six $S^{2-}$ anions. By sharing S(5) and S(8) with each other, Bi(1)$S_6$ octahedra propagate in the $b$-direction and form infinitely long chains, as shown in the first row of Figure 4. In the structure of $La_{0.92}Bi_{1.08}S_3$, the Bi(1)$S_4$ chains connect with the $Bi_2S_5$ blocks by sharing S(3) atoms. Bi(4) and La(1) disorderly distribute in the same site with occupancies of 0.253(7) and 0.747(7) respectively. Bi(4) is 7-coordinated to $S^{2-}$ anions while La(1) is 8-coordinated, as shown in the second and third row of Figure 4.

The coordination environments of La(2) and La(3) atoms is also shown in Figure 4. One can see all three $La^{3+}$ cations are 8-coordinated in a bicapped trigonal prismatic geometry. The La-S bonding distances range from 2.9906(16) to 3.136(4) Å for La(1), 2.9391(3) to 3.196(2) Å for La(2), and 2.8770(3) to 3.129(2) Å for La(3). The trigonal prisms also replicate along the *b*-axis and form LaS$_5$ infinite chains. The same coordination environment of La$^{3+}$ exists in La$_4$Bi$_2$S$_9$,[29] while in La$_7$Sb$_9$S$_{24}$[44] La$^{3+}$ cations are 9-coordinated to S$^{2-}$ anions in a tricapped trigonal prismatic geometry. The atom size of group 15 metal plays an important role in determining the coordination environment of La$^{3+}$ in La-group 15 metal-S systems. In fact, besides the 8-coordinated S$^{2-}$ anions, La(1), La(2) and La(3) neighbor to an extra S$^{2-}$ anion which is S(3), S(8) and S(7) respectively at a long distance. The distances are 4.8374(21) Å, 4.2041(28) Å and 3.5523(36) Å, respectively, which are not counted as bonds. This kind of long-distance neighboring causes larger atomic vibration along the associated direction. For instance, S(7) has a larger thermal parameter approximately along *a*-axis than in other directions due to the long distance neighboring between S(7) and La(3) approximately along *a*-axis. In comparison, the thermal parameters of Bi(4) are also relatively large but they are isotropic, which is related to the strong disorder of Bi(4).

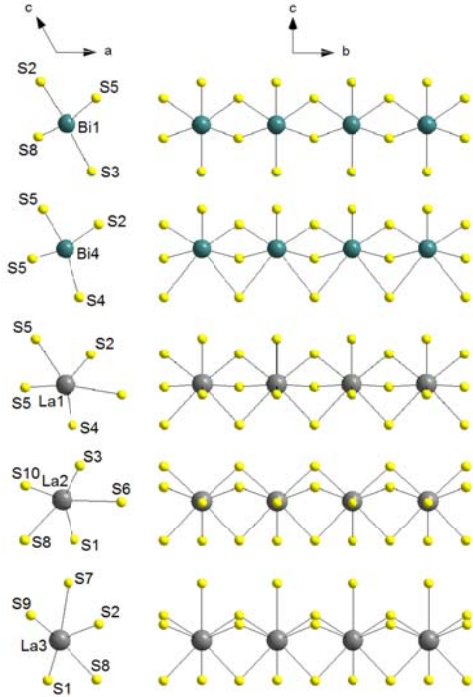

**Figure 4.** The Bi(1)S$_4$, Bi(4)S$_4$, La(1)S$_5$, La(2)S$_5$, and La(3)S$_5$ infinitely long chains viewed along the *b*-axis and the *a*-axis respectively.

**Heat Capacity.** The heat capacity measurements were performed from 2 to 300 K, as shown in Figure 5. No peaks or jumps associated with phase transitions were observed. At 300 K, the heat capacity approaches the value of 3*NR*, the so-called Dulong-Petit limit, where $N = 5$ (the number of atoms in the chemical formula) and $R = 8.314$ J mol$^{-1}$ K$^{-1}$ (the gas constant). This further proves there are no massive vacancies in the structure since the total heat capacity at 300 K is accurately contributed by 5$N_A$ ($N_A$ is Avogadro's number) atoms for 1 mol La$_{0.92}$Bi$_{1.08}$S$_3$.

**Optical Absorption.** Based on the charge balanced formula of (La$^{3+}$)$_{0.92}$(Bi$^{3+}$)$_{1.08}$(S$^{2-}$)$_3$ the compound is valence precise and is expected to be a semiconductor. Optical absorption spectra collected for La$_{0.92}$Bi$_{1.08}$S$_3$ at room temperature reveal a band gap transition near 1.0 eV, Figure 6, consistent with the black color of the crystals. For a direct optical transition the square of absorbance ($\alpha$) is expected to vary linearly with energy $\hbar\omega$ as $\alpha = (\hbar\omega - E_D)^{1/2}$ where $E_D$ is the direct optical band gap, and for an indirect transition the square root of $\alpha$ is expected to vary linearly with energy as $\alpha = (\hbar\omega - E_{ID})^2$ where $E_{ID}$ is the indirect band gap energy.[45] We have plotted the square and square root of absorption data as a function of energy (insets of Figure 6) and extracted a direct band gap component of 1.08(2) eV or an indirect band gap value of 0.98(2) eV. The linearity is better with the square of absorption data suggestive of a direct electronic transition. However, *ab-initio* theoretical calculations discussed below reveal an actually indirect band gap in La$_{0.92}$Bi$_{1.08}$S$_3$.

**Resistivity.** Temperature dependence of resistivity for a La$_{0.92}$Bi$_{1.08}$S$_3$ single crystal confirms the semiconducting nature of the compound. The resistivity is ~ 3×10$^6$ Ω·cm at 320 K and decreases gradually with increasing temperature, Figure 7. Furthermore, the temperature dependence of resistivity obeys thermally activated behavior as $\rho = \rho_o \exp(E_a/k_B T)$ where $\rho_o$ refers to a prefactor and $k_B$ is the Boltzmann's constant. This thermally activated behavior follows a linear trend when the ln$\rho$ vs 1/$T$ curve is plotted, as shown in the inset of Figure 8. Within the linear fitting process, the activation energy $E_a$ was extracted to be 0.36(1) eV. The activation energy is much smaller than the experimental band gap indicating the presence of mid-gap impurity energy levels which donate carriers to conduction bands or accept carriers from valence bands. The impurity energy levels may originate from the extrinsic defects like the existence of impurity atoms or intrinsic defects such as vacancies and dislocations in semiconductors.

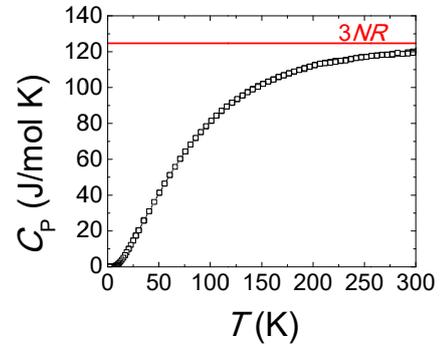

**Figure 5.** Heat capacity $C_P$ versus temperature for La$_{0.92}$Bi$_{1.08}$S$_3$. The heat capacity approaches 3*NR* at high temperatures.

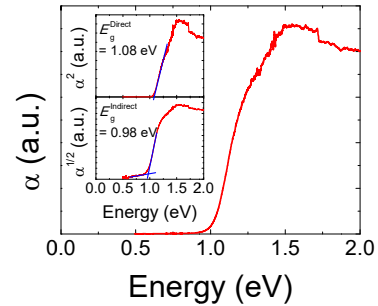

**Figure 6.** Absorption spectrum on La$_{0.92}$Bi$_{1.08}$S$_3$ measured at room temperature. The insets present the fits to direct band gap (top inset) and indirect band gap (bottom inset).

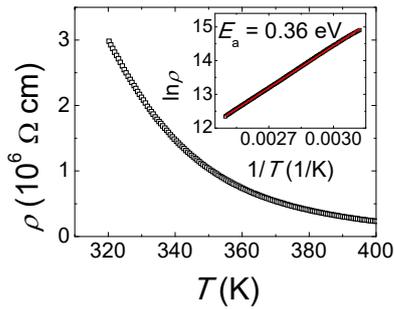

**Figure 7.** Electrical resistivity as a function of temperature for a La$_{0.92}$Bi$_{1.08}$S$_3$ single crystal. Inset: Arrhenius plot ln$\rho$ vs 1/$T$ showing a linear behavior for thermally activated conduction.

**Thermopower.** The thermopower of La$_{0.92}$Bi$_{1.08}$S$_3$ was measured on a pressed pellet from 300 K to 500 K. The sample is invariably *n*-type in the temperature region of 300-500 K and the Seebeck coefficient *S* is monotonically increasing from 50 μV/K to 350 μV/K, as shown in Figure 8. The *n*-type nature of La$_{0.92}$Bi$_{1.08}$S$_3$ indicates that electrons are the majority carriers in this material likely arising from a little bit sulfur vacancies. The magnitude of the thermopower of La$_{0.92}$Bi$_{1.08}$S$_3$ is comparable with that of the thermoelectric material Bi$_2$S$_3$ but the resistivity of La$_{0.92}$Bi$_{1.08}$S$_3$ is much higher,[46] which impedes the use of this material as a thermoelectric candidate.

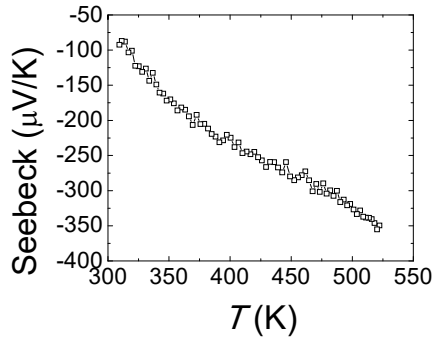

**Figure 8.** Thermopower as a function of temperature for a pressed pellet sample of La$_{0.92}$Bi$_{1.08}$S$_3$ showing *n*-type transport.

**Band Structure Calculations.** In Figure 9, we present the electronic band structure and density of states (DOS) of La$_{0.92}$Bi$_{1.08}$S$_3$. The band structure shows that La$_{0.92}$Bi$_{1.08}$S$_3$ is a semiconductor having an indirect gap of about 1.03 eV which is indicated with a red arrow. The bottom of the conduction band and the top of the valence band are very close between X and Γ points which could be the origin of the better linearity for the square of absorption data. This gap size is in good agreement with the experimental result of 0.98(2) eV. From the projected DOS on each atom in La$_{0.92}$Bi$_{1.08}$S$_3$, the energy range -2.0 eV to Fermi level is dominated by the S-*p* orbital states forming relatively narrow bands (low dispersion) while the La-*f* and Bi-*p* orbital states are located about 1.0 eV above the Fermi energy and form much wider bands.

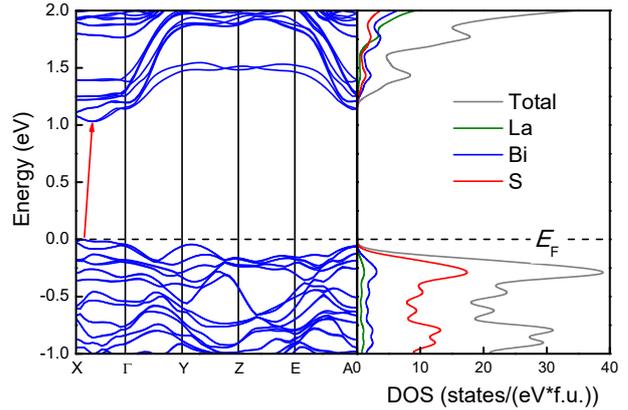

**Figure 9.** Band structures, total and partial DOS patterns of La$_{0.92}$Bi$_{1.08}$S$_3$. The Fermi energy is positioned at zero.

## CONCLUDING REMARKS

La$_{0.92}$Bi$_{1.08}$S$_3$ adopts a new tunnel-type structure. The phase appears to exhibit a certain phase width associated with the mixing of the trivalent La and Bi sites. The material is an anisotropic semiconductor with *n*-type charge transport and an indirect band gap of ~ 1 eV. The La$_{1-x}$Bi$_{1+x}$S$_3$ phase was sought for by Ecrepont *etal* who reported they were unable to definitively identify it and instead isolated La$_4$Bi$_2$S$_9$,[29] the only other known compound in the La-Bi-S system. The exact stoichiometric phase of LaBiS$_3$ may not be stable with a significant phase width of at least up to *x*=0.08 being necessary for stability.

## ASSOCIATED CONTENT

### Supporting Information

X-ray crystallographic data (CIF), tables for detailed structure parameters and Rietveld refinement of powder X-ray diffraction data are provided.

## AUTHOR INFORMATION

### Corresponding Author
*E-mail: m-kanatzidis@northwestern.edu

### Notes
The authors declare no competing financial interest.

## ACKNOWLEDGMENT

This work was supported by the U.S. Department of Energy, Office of Science, Basic Energy Sciences, Materials Sciences and Engineering Division. Use of the Center for Nanoscale Materials, including resources in the Electron Microscopy Center, was supported by the U. S. Department of Energy, Office of Science, Office of Basic Energy Sciences, under Contract No. DE-AC02-06CH11357. Work done at Nanjing University (electronic structure calculations by H. Liu and X. Wan) is supported by the NSF of China (Grant Nos. 11374137, and 11525417).

TOC (For Table of Contents Only)

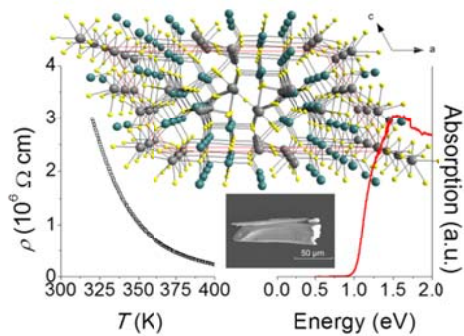

TOC synopsis

A new bismuth chalcogenide $La_{0.92}Bi_{1.08}S_3$ in the monoclinic structure has been discovered via solid state reaction. $La_{0.92}Bi_{1.08}S_3$ adopts a novel structure with positional disorder. This material shows a semiconducting behavior with a band gap of ~ 1 eV and activation energy for transport of 0.36(1) eV. Band structure calculations confirm the gap is indirect.